\documentclass[12pt]{article}

\usepackage[utf8]{inputenc}
\usepackage{latexsym, graphicx} 
\usepackage{amsmath,amsfonts,amssymb}
\usepackage{cancel}
\usepackage{mathrsfs}
\usepackage{tensor}
\usepackage{xcolor}
\usepackage{cite}
\usepackage[colorlinks=true,linkcolor=blue,citecolor=blue,urlcolor=blue,filecolor=black]{hyperref}

\renewenvironment{thebibliography}[1]{
  \begin{oldthebibliography}{#1}
    \setlength{\itemsep}{2pt}
    \setlength{\parskip}{2pt}
}
{
  \end{oldthebibliography}
}

\numberwithin{equation}{section} 
\usepackage{pict2e}
\makeatletter
\newcommand{\loplus}{\mathbin{\mathpalette\dog@lsemi{+}}}
\newcommand{\dog@lsemi}[2]{\dog@semi{#1}{#2}{270,90}}
\newcommand{\dog@semi}[3]{%
  \begingroup
  \sbox\z@{$\m@th#1#2$}%
  \setlength{\unitlength}{\dimexpr\ht\z@+\dp\z@\relax}%
  \makebox[\wd\z@]{\raisebox{-\dp\z@}{%
    \begin{picture}(1,1)
    \linethickness{\variable@rule{#1}}
    \roundcap
    \put(0.5,0.5){\makebox(0,0){\raisebox{\dp\z@}{$\m@th#1#2$}}}
    \put(0.5,0.5){\arc[#3]{0.5}}
    \end{picture}%
  }}%
  \endgroup
}
\newcommand{\variable@rule}[1]{%
  \fontdimen8  
  \ifx#1\displaystyle\textfont3\else
    \ifx#1\textstyle\textfont3\else
      \ifx#1\scriptstyle\scriptfont3\else
        \scriptscriptfont3\relax
  \fi\fi\fi
}
\makeatother

\newcommand*{\scri}{\ensuremath{\mathscr{I}}}
\newcommand*{\lied}{\mathop{}\!\mathcal{L}}
\newcommand*{\dd}{\mathop{}\!d}

\newcommand{\zbar}{\bar{z}}
\newcommand{\mM}{\mathcal{M}}
\newcommand{\tmM}{\tilde{\mathcal{M}}}
\newcommand{\tg}{\tilde{g}}
\newcommand{\Nvac}{N^{\text{vac}}}
\newcommand{\CFR}{\text{CFR}}

\newcommand*{\hateq}{\mathop{}\! \hat{=} \mathop{}\! }

\begin{document}

\begin{titlepage}
  \thispagestyle{empty}

  \begin{flushright}
  \end{flushright}

\vskip3cm

\begin{center}  
{\large\textbf{Schwarzian Transformations at Null Infinity}\\ \vskip 0.2cm
\textit{or}\, \textbf{The Unobservable Sector of Celestial Holography}}

\vskip1cm

\centerline{Kevin Nguyen}

\vskip1cm

{\it{Department of Mathematics, King's College London, London, United Kingdom}}\\
\vskip 0.5cm
{kevin.nguyen@kcl.ac.uk}

\end{center}

\vskip1cm

\begin{abstract}
I describe the Schwarzian behavior of the gravitational field at null infinity under superrotations, to be understood as an \textit{unobservable} gauge artifact. To this end I review the induced geometry of null infinity and the construction of the gauge-invariant News tensor that characterizes gravitational radiation. I discuss potentially important implications for the celestial holography program, suggesting in particular that the uplift of the AdS$_3$/CFT$_2$ correspondence directly relates to the unobservable gauge sector.
\end{abstract}

\end{titlepage}

{\hypersetup{linkcolor=black}
  \tableofcontents
  }

\section{Introduction}
The asymptotic structure of General Relativity has been the subject of renewed interest as it appears to encode important universal information about the infrared structure of scattering amplitudes in perturbative quantum gravity. Strominger made the key observation \cite{Strominger:2013jfa,Strominger:2014pwa,He:2014laa} that the leading soft graviton theorem of Weinberg \cite{Weinberg:1965nx} is actually nothing else than the Ward identity associated with the asymptotic symmetries discovered by Bondi, van der Burg, Metzner and Sachs (BMS) back in 1962 \cite{Bondi:1962px,Sachs:1962wk,Sachs:1962zza}. Further extensions of the BMS group of asymptotic symmetries have been proposed \cite{Barnich:2009se,Barnich:2010eb,Campiglia:2014yka,Campiglia:2015yka} and shown to relate directly to the subleading soft graviton theorems \cite{Campiglia:2014yka,Cachazo:2014fwa,Kapec:2014opa}. Other developments along these lines are reviewed in \cite{Strominger:2017zoo,Raclariu:2021zjz,Pasterski:2021rjz}.  

Asymptotically flat spacetimes admit a conformal compactification \textit{à la} Penrose \cite{Penrose:1962ij,Penrose:1965am}. Null rays and massless radiation reach part of the conformal boundary called \textit{null infinity}~$\scri$ with topology $\mathbb{S}^2 \times \mathbb{R}$. Interestingly Lorentz transformations act as SL$(2,\mathbb{C})$ global conformal transformations on the \textit{celestial sphere} $\mathbb{S}^2$, while translations have a more intricate nonlinear geometrical action. In order to fully exploit the asymptotic structure of gravity at $\scri$ and its implications for scattering amplitudes it turns out very useful to adopt a basis of boost eigenstates which are SL$(2,\mathbb{C})$ conformal primaries, rather than the more conventional momentum basis. As a result the $\mathcal{S}$-matrix can be recast as a set of correlation functions of a two-dimensional conformal field theory (CFT) living on the celestial sphere $\mathbb{S}^2$ \cite{He:2015zea}. The story gets even better~: the extended asymptotic symmetry algebra introduced by Barnich and Troessaert extends the global conformal symmetry to the full Virasoro symmetry \cite{Barnich:2009se,Barnich:2010eb}, such that standard CFT techniques potentially apply to this newly discovered \textit{celestial CFT}. In particular it contains a local stress tensor whose conformal Ward identity is the subleading soft graviton theorem \cite{Kapec:2016jld}. For more information and references about the celestial CFT program, the interested reader should consult \cite{Raclariu:2021zjz,Pasterski:2021rjz}.  

In parallel to these exciting developments emerged the idea that the celestial CFT is the right framework in which to investigate whether the holographic principle \cite{Susskind:1994vu} also applies to gravity with flat asymptotics. The program of \textit{celestial holography} aims at understanding quantum gravity from the celestial CFT, in a way analogous to the celebrated AdS/CFT correspondence \cite{Maldacena:1997re,Gubser:1998bc,Witten:1998qj}. One manifestation of this holographic principle that I find particularly striking is that the infrared soft factors of the gravitational $\mathcal{S}$-matrix \cite{Weinberg:1965nx} are fully governed by the correlation functions of a single primary operator \cite{Himwich:2020rro}. This primary operator is the Goldstone mode of spontaneously broken asymptotic supertranslation symmetries, and its celestial dynamics is fully encoded in a two-dimensional effective action that is derived by holographic means \cite{Nguyen:2020waf}. This is only one manifestation of celestial holography and much more work is needed to see how far the paradigm goes. One especially interesting approach to this problem is the proposed uplift of the AdS$_3$/CFT$_2$ correspondence \cite{deBoer:2003vf,Cheung:2016iub,Ball:2019atb} that offers the prospects to leverage much of the AdS/CFT technology to the benefits of celestial holography.

The present article serves two main purposes. The first is to offer a pedagogical review of the covariant approach to asymptotically flat gravity in terms of a null conformal boundary~$\scri$ and a News tensor that characterizes gravitational radiation. I will take this opportunity to connect this covariant approach to the coordinate-based Bondi--Sachs formalism and present a unified description of various important quantities involved in the construction of the physical News tensor, including the Geroch tensor \cite{Geroch1977}, the `vacuum' News tensor \cite{Compere:2018ylh,Campiglia:2020qvc}, the Liouville superboost field \cite{Compere:2018ylh}, etc. I will also describe the BMS symmetries and its various extensions from the perspective of null infinity $\scri$ and its induced conformal geometry. The second purpose is to critically discuss an apparent mismatch between the celestial CFT built out of the $\mathcal{S}$-matrix and the holographic CFT resulting from the uplifted AdS$_3$/CFT$_2$ correspondence. This mismatch concerns the value of the Virasoro central charge in the extended BMS algebra and the status of the respective local stress tensors. I conclude that the uplifted holographic stress tensor governs an \textit{unobservable} Schwarzian sector of asymptotically flat gravity and I speculate whether this sector could be that of \textit{infrared divergences}.    

\paragraph{Conventions.} A manifold $\tmM$ equipped with a lorentzian metric $\tg_{\alpha \beta}$ is called a \textit{spacetime}. The hatted equality sign $\hateq$ refers to an equality at null infinity $\scri$. Indices $\alpha,\beta,\gamma,...$ denote four-dimensional, $\mu,\nu,\rho,...$ denote three-dimensional, and $i,j,k,..$ denote two-dimensional coordinate indices, respectively.

\section{The geometry of null infinity}
\label{sec:geometry}
I start by recalling the definition of asymptotic flatness in relation to null infinity $\scri$ and the corresponding induced geometrical data. This approach largely relies on Penrose's conformal compactification \cite{Penrose:1962ij,Penrose:1965am}, which appears extremely well-suited to the description of massless fields and radiation infinitely far away from physical sources. I will mostly follow the treatment given by Geroch  \cite{Geroch1977}. Another useful reference is the review paper by Ashtekar \cite{Ashtekar:2014zsa}.\\ 
\newline
\textbf{Definition.} The physical spacetime $(\tmM,\tg_{\alpha \beta})$ is said to be \textit{asymptotically locally flat at null infinity} if there exists another spacetime $(\mM,g_{\alpha \beta})$ with boundary $\scri$ together with a smooth function $\Omega$ on $\mM$, such that
\begin{enumerate}
    \item $\tmM$ is diffeomorphic to $\mM-\scri$ (by which they are identified)\,,
    \item  on $\mM - \scri$ : $g_{\alpha \beta}=\Omega^2\, \tg_{\alpha \beta}$\,,
    \item at $\scri$ : $\Omega=0$,  $\nabla_\alpha \Omega\, \hat{\neq}\, 0$ and $\nabla^\alpha \Omega\, \nabla_\alpha \Omega \hateq 0$\,.
\end{enumerate}
The unphysical spacetime $(\mM, g_{\alpha \beta})$ is called an \textit{asymptote} of $(\tmM, \tg_{\alpha \beta})$. The first condition encodes the idea that $\mM$ is a (conformal) compactification of $\tmM$. The second condition together with the first part of the third condition states that the conformal boundary $\scri$ is infinitely far away with respect to the physical metric $\tg_{\alpha \beta}$. The condition $\nabla_\alpha \Omega\, \hat{\neq}\, 0$ ensures that $\Omega$ can be used as `radial' coordinate in a neighborhood of $\scri$, and identifies $n^\alpha \equiv g^{\alpha \beta} \nabla_\beta \Omega$ with the vector normal to $\scri$. Finally $n^\alpha n_\alpha \hateq 0$ states that $\scri$ is a null surface with respect to the unphysical metric and is therefore referred to as \textit{null infinity}. This last condition can actually be derived from Einstein's equations and some minimal assumptions on the falloff rate of the matter stress-energy tensor $\tilde T_{\alpha \beta}$ in a neighborhood of~$\scri$. To show this, we first write the relation between the physical and unphysical Einstein tensors,
\begin{equation}
\label{Gmunu conformal trans}
\tilde G_{\alpha \beta}=G_{\alpha \beta}+2 \Omega^{-1} \nabla_\alpha \nabla_\beta \Omega+g_{\alpha \beta}\left(3 \Omega^{-2} \nabla^\gamma \Omega\, \nabla_\gamma \Omega-2\Omega^{-1} \nabla^\gamma \nabla_\gamma \Omega \right)\,.
\end{equation}
Then we make the minimal assumption that $\tilde T_{\alpha \beta}$ admits a smooth limit to $\scri$, a condition easily satisfied by massless scalar fields and Maxwell fields for example \cite{Geroch1977}. Multiplying both sides of \eqref{Gmunu conformal trans} by $\Omega$, using Einstein's equations $\tilde G_{\alpha \beta}=8\pi G\, \tilde T_{\alpha \beta}$ and taking the limit $\Omega \to 0$, we conclude that the quantity 
\begin{equation}
f \equiv \Omega^{-1} \nabla^\alpha \Omega\, \nabla_\alpha \Omega\,,
\end{equation}
must also admit a smooth limit to $\scri$. In particular $n^\alpha n_\alpha \hateq 0$ such that the conformal boundary $\scri$ is a null surface.

The above definition is only concerned with local properties of null infinity. In particular any asymptote for which a portion of $\scri$ has been removed still satisfies this definition. It is then customary to require the global topology of $\scri$ to be $\mathbb{S}^2 \times \mathbb{R}$, in which case $(\tmM,\tg_{\alpha \beta})$ is said to be \textit{asymptotically flat at null infinity} \cite{AshtekarSchmidt,Ashtekar:2014zsa}. More elaborate definitions that ensure geodesic completeness of $\scri$ have also been given \cite{Penrose:1965am,Hawking:1973uf,GerochHorowitz,AshtekarSchmidt,Ashtekar:1981hw}. A global definition which further incorporates the spacetime structure at spatial infinity $i^0$ has been given by Ashtekar and Hansen \cite{Ashtekar:1978zz,Ashtekar1980}. See Wald's texbook for an overview of this subject \cite{Wald:1984rg}. In this paper I will not make use of these refined definitions. 

The definition of asymptotic flatness leaves significant ambiguity in the choice of conformal factor $\Omega$.
If $(\mM,g_{\alpha \beta},\Omega)$ satisfies the above criteria, so does $(\mM,\Omega',g'_{\alpha \beta})$ with
\begin{equation}
\label{conformal transformation}
\Omega' = \omega\, \Omega\,, \qquad g'_{\alpha \beta}=\omega^2 g_{\alpha \beta}\,,
\end{equation}
for $\omega$ any smooth and strictly positive function on $\mM$. Any sensible physical quantity should be independent of this choice, i.e., the Weyl rescaling \eqref{conformal transformation} should be considered a gauge redundancy. Said differently, \eqref{conformal transformation} provides an equivalence relation for the asymptotes. There is a unique equivalence class of asymptotes associated with a given asymptotically flat spacetime at null infinity (see theorem 2 in \cite{Geroch1977}). One finds that Weyl rescalings act like
\begin{subequations}
\begin{align}
n'^\alpha&=\omega^{-1} \left(n^\alpha + \Omega\, \nabla^\alpha \ln \omega \right)\,,\\
f'&=\omega^{-1} \left(f+2 n^\alpha \nabla_\alpha \ln \omega+\Omega\, \nabla_\alpha (\ln \omega)\, \nabla^\alpha (\ln \omega) \right)\,.
\end{align}
\end{subequations}
Using this gauge freedom, it is always possible to choose a conformal frame satisfying the \textit{Bondi condition} 
\begin{equation}
\label{Bondi frame}
f \hateq 0\,,
\end{equation} 
in which case Einstein's equations \eqref{Gmunu conformal trans} imply
\begin{equation}
\nabla_\alpha \nabla_\beta \Omega \hateq 0\,.
\end{equation}
As a direct consequence, we also have
\begin{equation}
\label{nabla n}
\nabla_\alpha\, n^\beta \hateq 0\,, \qquad \lied_n\, g_{\alpha \beta} \hateq 0\,.
\end{equation}
The condition \eqref{Bondi frame} leaves a residual rescaling gauge freedom parametrized by functions $\omega >0$ satisfying $\lied_n \omega = n^\alpha \nabla_\alpha \omega \hateq 0$.

We now have a closer look at the geometrical structure induced on $\scri$. Since the latter is a null surface, it is endowed with the \textit{Carrollian} structure $(q_{\mu \nu}, n^\mu)$ where $q_{\mu\nu}$ and $n^\mu$ are obtained by pullback of the unphysical metric $g_{\alpha \beta}$ and the normal vector $n^\alpha$, respectively. The induced metric $q_{\mu\nu}$ has signature $(0,+,+)$, and $n^\mu$ points along its degenerate direction, 
\begin{equation}
n^\mu q_{\mu \nu}=0\,.
\end{equation}
Note that under Weyl rescalings, these quantities transform like
\begin{equation}
\label{gauge rescaling}
q'_{\mu\nu}=\omega^2\, q_{\mu\nu}\,, \qquad n'^\mu = \omega^{-1}\, n^\mu\,.
\end{equation}
When the Bondi condition \eqref{Bondi frame} is satisfied, the Levi-Civita derivative operator $\nabla_\alpha$ induces a torsionfree and metric compatible derivative operator $D_\mu$ at $\scri$ satisfying
\begin{equation}
\label{Dmu}
D_\rho\, q_{\mu\nu}=0\,, \qquad D_\rho\, n^\mu=0\,.
\end{equation}
The proof of these statements, together with a detailed discussion of the underlying Carrollian affine connection, are given in appendix~\ref{app:connection}. In contrast to the case where the metric is non-degenerate, these conditions do not uniquely fix the induced connection. In the present context the connection coefficients left undetermined by \eqref{Dmu} actually encode non-universal information about gravitational radiation \cite{Geroch1977,Ashtekar:1981hw}.  For concreteness it is sometimes useful to introduce an adapted coordinate system $x^\mu=(u,x^i)$ in which the metric $q_{\mu\nu}$ takes the form
\begin{equation}
\label{u,x coordinates}
ds^2=0 \dd u^2 +q_{ij} \dd x^i \dd x^j\,, \qquad \partial_u q_{ij}=0\,.
\end{equation}
Here $x^i$ are the coordinates covering a `cut' of $\scri$ with topology of the sphere $\mathbb{S}^2$, and $q_{ij}$ is the induced two-dimensional euclidean metric. In this coordinate system $\Gamma^k_{ij}$ and $\Gamma^u_{ij}$ are the only nonzero Christoffel symbols, where the first are the Levi-Civita coefficients associated with the metric $q_{ij}$ while the second are left undetermined by \eqref{Dmu}. As shown explicitly in section~\ref{sec:Bondi} these appear to encode the shear tensor,  $\Gamma^u_{ij} \sim C_{ij}$, that describes gravitational waves passing through $\scri$. 

\section{News and Geroch tensors}
\label{sec:News}
After having described the geometry of null infinity, I turn to the quantities needed to properly characterize gravitational radiation. As is well-known, most of the physical information carried by gravitational waves is encoded in the \textit{News tensor} of Bondi and Sachs~\cite{Bondi:1962rkt,Bondi:1962px,Sachs:1962wk}. In the formalism used here, its construction was given by Geroch \cite{Geroch1977}.

It turns out that gravitational waves reaching $\scri$ can essentially be described in terms of the unphysical Ricci tensor, or more precisely in terms of the unphysical Schouten tensor
\begin{align}
S_{\alpha \beta}&\equiv R_{\alpha \beta}-\frac{1}{6} R\, g_{\alpha \beta}\,.
\end{align}
Indeed its projection $S_{\mu\nu}$ to $\scri$ acts as a potential for the leading order Weyl tensor \cite{Geroch1977}. It also satisfies the properties
\begin{equation}
S_{\mu\nu}\, n^\nu=0\,, \qquad S_{\mu\nu}\, q^{\mu\nu}=\mathcal{R}\,,
\end{equation}
where $\mathcal{R}$ is the scalar curvature of $D_\mu$ and $q^{\mu\nu}$ is any covariant tensor satisfying $q_{\mu \rho} q^{\rho\sigma} q_{\sigma \nu}=q_{\mu\nu}$. 
It is however not gauge invariant, since it transforms under \eqref{conformal transformation} like
\begin{align}
S'_{\mu\nu}&=S_{\mu \nu}-2 \omega^{-1} D_\mu D_\nu \omega+4 \omega^{-2} D_\mu  \omega\, D_\nu \omega- q_{\mu \nu}\, \omega^{-2} D_\rho \omega\, D^\rho \omega\,.
\end{align}
Fortunately one can construct a gauge-invariant tensor at $\scri$ in the elegant following way. Geroch proved that there exists a unique `kinematical' tensor at $\scri$, i.e., constructed out of universal geometrical data, that satisfies\footnote{In the adapted coordinates $(u,x^i)$ introduced in \eqref{u,x coordinates}, the Geroch tensor is simply the lift of a tensor $\rho_{ij}$ on the sphere $\mathbb{S}^2$ satisfying 
\begin{equation}
\rho_{[ij]}=0\,, \qquad  \rho_{ij}\, q^{ij}=\mathcal{R}\,, \qquad D_{[i} \rho_{j]k}=0\,,
\end{equation}
where $D_i$ and $\mathcal{R}$ are now the Levi-Civita connection and curvature associated with the two-dimensional euclidean metric $q_{ij}$.} \cite{Geroch1977} 
\begin{align}
\label{Geroch conditions}
\rho_{[\mu\nu]}=0\,, \qquad \rho_{\mu\nu}\, n^\nu=0\,, \qquad \rho_{\mu \nu}\, q^{\mu\nu}=\mathcal{R}\,, \qquad D_{[\rho} \rho_{\mu]\nu}=0\,.
\end{align}
The value of the Geroch tensor lies in its transformation under Weyl rescalings that is identical to that of $S_{\mu\nu}$,
\begin{subequations}
\label{rho weyl transf}
\begin{align}
\rho'_{\mu \nu}& = \rho_{\mu \nu}-2 \omega^{-1} D_\mu D_\nu \omega+4 \omega^{-2} D_\mu  \omega\, D_\nu \omega- q_{\mu \nu}\, \omega^{-2} D_\rho \omega\, D^\rho \omega\\
&=\rho_{\mu\nu}-2 \omega^{-1} D'_\mu D'_\nu \omega+ q'_{\mu\nu}\, \omega^{-2} D'_\rho \omega\, D'^\rho \omega\,.
\end{align}
\end{subequations}
This allows to define the gauge-invariant News tensor
\begin{equation}
N_{\mu\nu}\equiv \rho_{\mu\nu}-S_{\mu\nu}\,, \qquad N_{\mu\nu}\, n^\nu=N_{\mu\nu}\, q^{\mu\nu}=0\,.
\end{equation}
This is the physical quantity that characterizes gravitational radiation at $\scri$. The change in relative sign compared to the definition in \cite{Geroch1977} is chosen for consistency with the Bondi--Sachs formalism discussed in section~\ref{sec:Bondi}.

It is actually possible to explictly construct the Geroch tensor. In a conformal frame where $\mathcal{R}=\mathcal{R}^0$ is constant, the unique solution to \eqref{Geroch conditions} can be easily seen to be
\begin{equation}
\rho_{\mu\nu}^{0}=\frac{1}{2} \mathcal{R}^{0}\, q^{0}_{\mu\nu}\,,
\end{equation}
in which case the News is simply the tracefree part of $S_{\mu\nu}$. To find its expression in a more general frame still satisfying the Bondi condition \eqref{Bondi frame},
\begin{equation}
\label{q0}
q_{\mu\nu}=\omega^2\, q^0_{\mu\nu}=e^{2\psi} q^0_{\mu\nu}\,, \qquad \lied_n \omega=\lied_n \psi=0\,,
\end{equation}
we simply need to use the second equation of \eqref{rho weyl transf} together with the Weyl transformation of the curvature scalar,
\begin{equation}
\mathcal{R}=\omega^2 \mathcal{R}'+2 \omega D'_\rho D'^\rho \omega-2 D'_\rho \omega\, D'^\rho \omega\,.
\end{equation}
This allows us to write
\begin{equation}
\label{rho=R-T}
\rho_{\mu\nu}=\frac{1}{2} \mathcal{R}\, q_{\mu\nu}-\mathcal{T}_{\mu\nu}\,,
\end{equation}
where the traceless tensor $\mathcal{T}_{\mu\nu}$ is given by
\begin{align}
\label{Tmunu}
\mathcal{T}_{\mu\nu}&=2 \left[ \omega^{-1} D_\mu D_\nu \omega \right]^{\text{TF}}=2\left[D_\mu D_\nu \psi+D_\mu \psi D_\nu \psi \right]^{\text{TF}}\,,
\end{align}
in agreement with the expression given in appendix~B of \cite{Campiglia:2020qvc}. The News tensor thus admits the alternative expression
\begin{equation}
\label{News rewriting}
N_{\mu\nu}=-[S_{\mu\nu}]^{\text{TF}}-\mathcal{T}_{\mu\nu}\,.
\end{equation}

\textit{Remark :} Working again in adapted coordinates $(u,x^i)$, one can introduce the \textit{superboost Liouville field} $\Phi$ of Compère, Fiorucci and Ruzziconi \cite{Compere:2018ylh} by fixing the reference metric to be (the lift of) the flat metric $q^0_{ij}=\delta_{ij}$ and defining $\Phi=-\frac{1}{2} \psi$, so that
\begin{align}
\label{Liouville equation}
\mathcal{R}&=D_i D^i \Phi\,,
\end{align}
and 
\begin{equation}
\Nvac_{ij}\equiv \mathcal{T}_{ij}=\left[\frac{1}{2} D_i \Phi D_j \Phi-D_i D_j \Phi\right]^{\text{TF}}\,.
\end{equation}
These authors introduce the additional notations
\begin{equation}
\hat{N}_{ij}^{(\CFR)}=N_{ij}\,, \qquad N^{(\CFR)}_{ij}=-[S_{ij}]^{\text{TF}}\,,
\end{equation}
such that \eqref{News rewriting} takes the form
\begin{equation}
\hat{N}^{(\CFR)}_{ij}=N^{(\CFR)}_{ij}-\Nvac_{ij}\,.
\end{equation}
The `vacuum News'\footnote{This terminology is unfortunate since the physical (gauge-invariant) News $N_{ij}$ is identically zero in absence of radiation by contradistinction with $\Nvac_{ij}$. I prefer to refer to $\Nvac_{ij}=\mathcal{T}_{ij}$ as the tracefree Geroch tensor.} $\mathcal{T}_{ij}$ interestingly coincides with the tracefree part of the stress tensor of a Liouville theory with action
\begin{equation}
S_{\text{Liouville}}=\int d^2x\, \sqrt{q} \left(\frac{1}{2} D^i \Phi D_i \Phi+\mathcal{R}\, \Phi \right)\,,
\end{equation}
while \eqref{Liouville equation} is the corresponding equation of motion.
It is well-known that the stress tensor of any CFT$_2$ is universally described by this action (see \cite{Nguyen:2020lbg} for a review), which perhaps hints at a role played by the above quantities in \textit{celestial holography}. In particular when complex stereographic coordinates $x^i=(z,\zbar)$ are used to cover the sphere~$\mathbb{S}^2$, it can be shown that $\mathcal{T}_{zz}$ reduces to a Schwarzian derivative \cite{Compere:2016jwb}. I will come back to these points in section~\ref{sec:Schwarzian}.

\section{Asymptotic symmetries}
\label{sec:symmetries}
Asymptotic symmetries of asymptotically flat spacetimes play a prominent role in our modern understanding of perturbative quantum gravity, and one should also expect them to underlie the basic structure of celestial holography.  I describe below the three common proposals for what the symmetries of gravity with flat asymptotics are, from the perspective of the universal geometry induced at null infinity. 

\paragraph{Global BMS algebra.} The original discovery that asymptotically flat spacetimes possess asymptotic symmetries was made by Bondi, van der Burg, Metzner and Sachs \cite{Bondi:1962px,Sachs:1962wk,Sachs:1962zza}. The corresponding symmetry group is called nowadays the \textit{global BMS group}.
The BMS symmetries can be viewed as the subgroup of diffeomorphisms of $\scri$ which preserve the pair $(q_{\mu\nu},n^\mu)$ up to a Weyl rescaling \eqref{gauge rescaling}, i.e., up to a gauge transformation.\footnote{This is also referred to as the conformal Carroll algebra of level 2,  $\mathfrak{bms}=\mathfrak{ccarr}_2$ \cite{Duval:2014uva,Duval:2014lpa}.} Infinitesimally, they are therefore generated by vector fields $\xi^\mu$ satisfying
\begin{equation}
\label{infinitesimal BMS}
\lied_\xi q_{\mu\nu}=2 \kappa\, q_{\mu\nu}\,, \qquad \lied_\xi n^\mu=-\kappa\, n^\mu\,.
\end{equation}
They must also preserve the Bondi condition $\lied_n q_{\mu\nu}=0$ which requires the function $\kappa$ to satisfy $\lied_n \kappa=0$. This can be shown from the following equality,
\begin{equation}
0 \stackrel{!}{=}  \lied_\xi \lied_n q_{\mu\nu}=\lied_n \lied_\xi q_{\mu\nu}+ \lied_{[\xi,n]} q_{\mu\nu}=2 q_{\mu\nu} \lied_n \kappa-\lied_{\kappa n} q_{\mu\nu}=2 q_{\mu\nu} \lied_n \kappa\,.
\end{equation}
A subset of these vector fields are of the simple form
\begin{equation}
\label{supertranslation}
\xi^\mu= f\, n^\mu\,, \qquad \lied_n f=0\,,
\end{equation}
satisfying the above constraints with $\kappa=0$. They form the abelian subalgebra $\mathfrak{s} \subset \mathfrak{bms}$ of \textit{supertranslations}. One can show that the Lie bracket of a supertranslation generator $f\, n^\mu \in \mathfrak{s}$ with a generic vector field $\xi^\mu \in \mathfrak{bms}$ is again a supertranslation generator,
\begin{align}
\label{ideal}
\xi'^{\mu}\equiv [\xi, f n]^\mu=\kappa_{\xi'}\, n^\mu\,, \qquad \kappa_{\xi'}=\left(\lied_\xi f-\kappa_\xi f \right)\,, \qquad \lied_n \kappa_{\xi'}=0\,.
\end{align}
Thus $\mathfrak{s}$ is also an ideal of the $\mathfrak{bms}$ Lie algebra. The structure of the quotient $\mathfrak{bms}/\mathfrak{s}$ can be understood in the following way. For any $\xi^\mu \in \mathfrak{bms}/\mathfrak{s}$, we adopt the parametrization
\begin{equation}
\xi^\mu=\bar \xi^\mu+\alpha\, n^\mu\,, \qquad \lied_n \bar \xi^\mu=0\,,
\end{equation}
and we find that the second equation of \eqref{infinitesimal BMS} requires
\begin{equation}
\lied_n \alpha=\kappa\,.
\end{equation}
Lowering with the degenerate metric, $\bar \xi_\mu=q_{\mu\nu}\, \bar \xi^\nu$, we find the constraints
\begin{align}
\label{Lorentz}
\bar \xi_\mu\, n^\mu=0\,, \qquad \lied_n \bar \xi_\mu=0\,, \qquad D_\mu \bar \xi_\nu +D_\nu \bar \xi_\mu=2 \kappa\, q_{\mu\nu}\,.
\end{align}
This should be understood as the lift to $\scri$ of the two-dimensional conformal Killing equation, which is more explicit in adapted coordinates $x^\mu=(u,x^i)$. As is well-known the \textit{globally well-defined} solutions to \eqref{Lorentz} form a $\mathfrak{sl}(2,\mathbb{C})=\mathfrak{so}(3,1)$ Lie algebra. Putting this together, the BMS Lie algebra has the semi-direct sum structure
\begin{equation}
\mathfrak{bms}=\mathfrak{so}(3,1) \loplus \mathfrak{s}\,.
\end{equation}
This is exactly like the Poincaré algebra, except that the translation algebra $\mathfrak{t}=\mathbb{R}^{3,1}$ is replaced by the infinite-dimensional abelian algebra $\mathfrak{s}$. This is not a coincidence as it is understood that Poincaré transformations of Minkowski space induce an action at $\scri$ which is precisely of the type described above \cite{Geroch1977}. The generators of the translation subalgebra $\mathfrak{t} \subset \mathfrak{s}$ can be isolated by the following conditions \cite{Geroch1977},
\begin{align}
\label{translations}
\left(D_\mu D_\nu +\frac{1}{2} \rho_{\mu\nu}\right) f \propto q_{\mu\nu}\,,
\end{align}
where $\rho_{\mu\nu}$ is the Geroch tensor defined in \eqref{Geroch conditions}. In a conformal frame where $q_{\mu\nu}$ is the unit round sphere metric, the solutions to \eqref{translations} are the four lowest spherical harmonics \cite{Ashtekar:2014zsa}.

To make the symmetry algebra fully explicit, we can again use adapted coordinates $x^\mu=(u,x^i)$ such that we can write a generic symmetry generator in the form
\begin{equation}
\xi^\mu=\left(f+\frac{u}{2} D_j \xi^j\, ,\, \xi^i\right)\,, \qquad \partial_u f=\partial_u \xi^i=0\,,
\end{equation}
where $\xi^i$ is a two-dimensional conformal Killing vector field satisfying
\begin{equation}
\label{Killing equation}
D_i \xi_j+D_j \xi_i=D_k \xi^k\, q_{ij}\,.
\end{equation}
Then the $\mathfrak{bms}$ algebra takes the familiar form
\begin{align}
\label{bms algebra}
\hat{\xi}_{12}^\mu= [\xi_1,\xi_2]^\mu\,,
\end{align}
with
\begin{subequations}
\label{bms algebra bis}
\begin{align}
f_{12}&=\xi_1^i D_i f_2+\frac{1}{2}f_1 D_i \xi_2^i\ - (1 \leftrightarrow 2)\,,\\
\xi^i_{12}&=\xi_1^j D_j \xi_2^i-(1 \leftrightarrow 2)\,.
\end{align}
\end{subequations}

\paragraph{Extended BMS algebra.} An extension of the BMS algebra was proposed more recently by Barnich and Troessaert \cite{Barnich:2009se,Barnich:2010eb}. The proposal is simply to consider \textit{local} solutions of the conformal Killing equation \eqref{Killing equation}. In complex stereographic coordinates $x^i=(z,\zbar)$ covering the sphere $\mathbb{S}^2$, the corresponding vector fields have (anti)-meromorphic components $\xi^z(z)$ and $\xi^{\zbar}(\zbar)$. Therefore the extended BMS algebra takes the form
\begin{equation}
\mathfrak{bms^e}=\left[ \mathfrak{diff}(\mathbb{S}^1) \oplus \mathfrak{diff}(\mathbb{S}^1)  \right] \loplus \mathfrak{s}^*\,.
\end{equation}
Note that $\mathfrak{diff}(\mathbb{S}^1)$ is also the centerless Virasoro algebra and it is often this latter terminology that is used. The \textit{superrotation} vector fields sitting in the quotient $\left[ \mathfrak{diff}(\mathbb{S}^1) \oplus \mathfrak{diff}(\mathbb{S}^1)  \right]/\mathfrak{so}(3,1)$ are not globally well-defined since they have poles at isolated points on the Riemann sphere~$\mathbb{S}^2$. It might therefore look like these additional transformations should not be allowed if we restrict to everywhere smooth metric fields $q_{ij}$ and cuts of $\scri$ with sphere topology.\footnote{Situations where cuts of $\scri$ are more general Riemann surfaces have been considered in \cite{Foster1978,Foster1987,Barnich:2021dta}. The singular Virasoro superrotations have also been interpreted as inserting cosmic string defects in the bulk of spacetime \cite{Strominger:2016wns} or defects on the celestial sphere \cite{Adjei:2019tuj}.} But the situation is identical to that of two-dimensional CFTs, and the existence of these non-global symmetries is sufficient to guarantee the existence of locally conserved currents \cite{Belavin:1984vu}. This justifies the extended version of the BMS algebra. Finally the supertranslation algebra $\mathfrak{s}^*$ is generated by functions $f$ that can also have singularities on the sphere in order for the Lie algebra \eqref{bms algebra bis} to close, and it is therefore larger than the algebra $\mathfrak{s}$ of smooth supertranslations. 

\paragraph{Generalized BMS algebra.} A generalization of the asymptotic symmetry algebra has been proposed by Campiglia and Laddha \cite{Campiglia:2014yka,Campiglia:2015yka}. Generalized BMS symmetries preserve the conformal class $(\epsilon_{\mu\nu\rho},n^\mu)$ where $\epsilon_{\mu\nu\rho}$ is the volume form at $\scri$, together with the Bondi condition $\lied_n q_{\mu\nu}=0$. To show this, I first introduce a degenerate tetrad $e_\mu^i\big|_{i=1,2}$ satisfying
\begin{equation}
q_{\mu\nu}=\delta_{ij}\, e^i_\mu e^j_\nu\,, \qquad e^i_\mu n^\mu=0\,.
\end{equation}
Here $\delta_{ij}$ can be understood to be the euclidean metric on the tangent space of the sphere $\mathbb{S}^2$. In order to construct the volume form $\epsilon_{\mu\nu\rho}$ we need a third linearly independent one-form $l_\mu$ normalized with $l_\mu n^\mu=1$, such that
\begin{equation}
\epsilon_{\mu\nu\rho}=\frac{1}{3!} l_{[\mu} e^1_\nu e^2_{\rho]}\,.
\end{equation}
Note that the ambiguity $l_\mu \to l_\mu + h\, e^i_\mu$ in defining $l_\mu$ does not affect the volume form. Finally we deduce the Weyl transformations of $e^i_\mu$ and $l_\mu$ from that of $q_{\mu\nu}$ and $n^\mu$,
\begin{equation}
e'^i_\mu=\omega\, e^i_\mu\,, \qquad l'_\mu=\omega\, l_\mu\,,
\end{equation}
so that
\begin{equation}
\epsilon'_{\mu\nu\rho}=\omega^3\, \epsilon_{\mu\nu\rho}\,.
\end{equation}
Infinitesimally, generalized BMS symmetries are thus generated by vector fields satisfying
\begin{equation}
\label{generalized BMS}
\lied_\xi \epsilon_{\mu\nu\rho}=3 \kappa\, \epsilon_{\mu\nu\rho}\,, \qquad \lied_\xi n^\mu=-\kappa\, n^\mu\,, \qquad \lied_n \kappa=0\,,
\end{equation}
where the last equation again follows from the Bondi condition. The corresponding Lie algebra is very similar to that of the global $\mathfrak{bms}$ algebra. In particular, there is still an abelian ideal $\mathfrak{s}$ of supertranslations characterized by \eqref{supertranslation}-\eqref{ideal}. The only difference stems from the absence of the conformal Killing equation constraint, such that the quotient $\mathfrak{bms^g}/\mathfrak{s}$ now contains the generators of \textit{all} smooths diffeomorphisms of the sphere,
\begin{equation}
\mathfrak{bms^g}=\mathfrak{diff}(\mathbb{S}^2) \loplus \mathfrak{s}\,.
\end{equation}
In adapted coordinates $(u,x^i)$, the explicit form of the Lie algebra \eqref{bms algebra}-\eqref{bms algebra bis} still holds. Note that the first equation of \eqref{generalized BMS} is simply used to eliminate $\kappa$ in terms of $\xi^\mu$,
\begin{equation}
\kappa=\frac{1}{3} D_\mu \xi^\mu=\frac{1}{3} \left(D_i \xi^i+\partial_u \xi^u \right)=\frac{1}{2} D_i \xi^i\,,
\end{equation}
but does not actually impose any additional constraint on $\xi^\mu$.\\

\textit{Remark :} In the Bondi--Sachs formalism, asymptotic symmetries are realized as bulk diffeomorphisms having support at $r \to \infty$ and preserving the Bondi gauge \eqref{physical metric}. There are bulk diffeomorphisms involving reparametrizations of the radial coordinate $r$ which precisely act like Weyl rescalings at $\scri$ \cite{Barnich:2011ty,Barnich:2010eb,Barnich:2016lyg}. It is customary to use these additional diffeomorphisms in order to undo the Weyl rescaling induced in \eqref{infinitesimal BMS} for example. Thus when $q_{ij}$ is kept fixed one obtains the extended BMS alegbra, while if only $\sqrt{q}$ is kept fixed one obtains the generalized BMS algebra.

\section{Superrotations and Schwarzian transformations}
\label{sec:Schwarzian}
We are now ready to show that the Geroch tensor $\rho_{ij}$ behaves like the stress tensor of a CFT$_2$, and more specifically that it has an anomalous Schwarzian transformation under Virasoro superrotations.

We start by choosing the reference metric $q^0_{\mu\nu}$ in \eqref{q0} to be the flat metric on the Riemann sphere covered by complex stereographic coordinates $(z',\zbar')$,
\begin{equation}
q^0_{ij} \dd x'^i \dd x'^j=\dd z' \dd \zbar'\,.
\end{equation}
Then we consider the \textit{conformal symmetry}, consisting of a meromorphic change of coordinates
\begin{equation}
z'=\Pi(z)\,, \qquad \zbar'=\zbar\,,
\end{equation}
followed by the Weyl rescaling
\begin{equation}
\label{Weyl q0}
q^0_{ij} \quad \mapsto \quad q_{ij}=(\partial_z \Pi)^{-1}\, q^0_{ij}\,, 
\end{equation}
so that the total transformation is a symmetry of the background metric (except at isolated points)
\begin{equation}
(ds^0)^2=\dd z' \dd \zbar'=\partial_z \Pi \dd z \dd \zbar \quad \mapsto \quad ds^2= \dd z \dd \zbar\,.
\end{equation}
From \eqref{rho=R-T}-\eqref{Tmunu} we find that the component $\rho_{zz}$ of the Geroch tensor takes the familiar form of a Schwarzian derivative,
\begin{equation}
\label{Schwarzian}
\rho_{zz}=-\mathcal{T}_{zz}=\frac{\partial_z^3 \Pi}{\partial_z \Pi}-\frac{3}{2}\left(\frac{\partial_z^2 \Pi}{\partial_z \Pi} \right)^2\equiv S[\Pi(z);z]\,.
\end{equation}
The overall coefficient which would be the central charge if $\rho_{zz}$ really was the stress tensor of a two-dimensional CFT, is simply unity. As usual this Schwarzian transformation can be traced back to the Weyl transformation \eqref{Weyl q0} of the background metric, and although it might be tempting to view \eqref{Schwarzian} as a manifestation of a celestial CFT, we should remember that the Geroch tensor was introduced precisely such that the physical News be invariant under Weyl rescalings. Therefore the Schwarzian transformation \eqref{Schwarzian} is pure gauge and unobservable.\\

\textit{Remark :} It is customary to fix a conformal frame where the metric $q_{ij}$ is the unit round sphere metric rather than the flat metric. However this does not affect the end result \eqref{Schwarzian}. To show this one simply has to modify the Weyl rescaling \eqref{Weyl q0} by
\begin{equation}
q^0_{ij} \quad \mapsto \quad q_{ij}=\gamma^2\, (\partial_z \Pi)^{-1}\, q^0_{ij}\,, \qquad \gamma=\frac{2}{1+z \zbar}\,,
\end{equation}
and again make use of \eqref{rho=R-T}-\eqref{Tmunu} together with the nonzero Christoffel symbol $\Gamma^z_{zz}=2 \partial_z \ln \gamma$. 

\section{Relation with the Bondi--Sachs formalism}
\label{sec:Bondi}
So far everything has been worked out in terms of the geometry of null infinity $\scri$, and it might be useful to make connection with the coordinate-based approach that is widely used in the literature. In the latter one writes the physical metric in Bondi gauge (see \cite{Compere:2018aar} for a review), 
\begin{align}
\label{physical metric}
\dd\tilde s^2=&-\dd u^2-2 \dd u \dd r+ \left(r^2 q_{ij}+ r\, C_{ij} \right) \dd x^i \dd x^j+...\,,
\end{align}
where $...$ refer to terms that are subleading in a large-$r$ expansion. Here $q_{ij}$ is a metric on the sphere $\mathbb{S}^2$ while the shear $C_{ij}$ is a traceless symmetric tensor. 

Now we construct the unphysical metric following section~\ref{sec:geometry}. We make the choice of conformal factor $\Omega=1/r$, such that the unphysical metric is
\begin{align}
\label{unphysical metric}
\dd s^2=&- \Omega^2 \dd u^2+2 \dd u \dd \Omega+\left(q_{ij}+\Omega\, C_{ij} \right) \dd x^i \dd x^j+...\,.
\end{align}
The normal vector is simply
\begin{equation}
n_\alpha=\delta^\Omega_\alpha\,, \qquad n^\alpha=\delta^\alpha_u+O(\Omega^2)\,,
\end{equation}
and the Bondi condition \eqref{Bondi frame} is therefore satisfied.

Null infinity $\scri$ is the surface at $\Omega=0$ and is covered with the adapted coordinates $x^\mu=(u,x^i)$. We need to split the tangent bundle $T \mathcal{M}$ at $\scri$ into $T \scri$ and its complement, for which there is no canonical procedure as explained in appendix~\ref{app:connection}. We will do this by simply discarding the $\Omega$-components of the various tensors and their covariant derivatives. By explicit computation, we find that the only nonzero induced connection coefficients at~$\scri$ are
\begin{equation}
\label{Gamma_u}
\Gamma^k_{ij}=\frac{1}{2}q^{kl}\left(\partial_i q_{lj}+\partial_j q_{il}-\partial_l q_{ij} \right)\,, \qquad \Gamma^u_{ij}=-\frac{1}{2} C_{ij}\,.
\end{equation}
This is what we expected from appendix~\ref{app:connection}. The spatial components of the induced connection are the Levi-Civita coefficients associated with the spatial metric $q_{ij}$, while $\Gamma^u_{ij}$ is directly proportional to the shear and therefore encodes information about gravitational radiation. We also compute the unphysical Schouten tensor, and we find
\begin{equation}
S_{ij}=q_{ij}-\partial_u C_{ij}\,, \qquad S_{uu}=S_{ui}=0\,, \qquad (\Omega=0)\,.
\end{equation}
In case that $q_{ij}=q^0_{ij}$ is the metric of the unit round sphere (without punctures), the physical News tensor \eqref{News rewriting} would therefore simply be
\begin{equation}
N_{ij}=-[S_{ij}]^{\text{TF}}=\partial_u C_{ij}\,.
\end{equation}
However, one should be careful not to take this formula outside of its regime of validity. In general the Geroch tensor is explicitly needed.

Minkowski spacetime is a vacuum of General Relativity since it has zero energy among other things. It is however not invariant under BMS symmetries and there is in fact an infinite family of degenerate vacua. Compère and Long have explicitly constructed these vacua by performing finite extended BMS transformations from Minkowski space \cite{Compere:2016jwb} (the case with generalized BMS transformations has been treated in \cite{Compere:2018ylh}). Adopting a frame where the metric $q_{ij}$ is the unit round sphere metric in complex stereographic coordinates, $q_{ij} \dd x^i \dd x^j=\gamma^2 \dd z \dd \zbar$, the vacuum solutions take the form \eqref{physical metric} with
\begin{align}
C_{zz}=-\left(u+f(z,\zbar)\right) S[\Pi(z);z]-2\, D^2_z f(z,\zbar)=\left(u+f(z,\zbar)\right) \mathcal{T}_{zz}-2\, D^2_z f(z,\zbar)\,.
\end{align}
Here $f(z,\zbar)$ and $\Pi(z)$ are the parameters of the supertranslation and Virasoro superrotation transformations, respectively. Again we see the appearance of a Schwarzian derivative associated with the Virasoro part of the transformation. But this quantity is pure gauge and is cancelled out in the News thanks to the Geroch tensor \eqref{Schwarzian},
\begin{equation}
N_{zz}=\partial_u C_{zz}+\rho_{zz}=0\,.
\end{equation}
This is reassuring as no gravitational waves have been created by acting with a symmetry transformation on empty Minkowski space.

\section{Discussion}
\label{sec:discussion}

\paragraph{Extended vs.~generalized BMS.} I want to emphasize the important conceptual difference between the  extended and generalized BMS transformations. Indeed the former leaves the metric $q_{ij}$ invariant up to a gauge transformation, by contrast to the latter one. Hence generalized BMS transformations deserve to be called symmetries
if and only if the (conformal class of) metric~$q_{ij}$ is a genuine \textit{dynamical} field rather than a fixed background structure.
\begin{quote}
\centering
\textit{Do we need to fix $q_{ij}$ as part of the boundary conditions\\ in order to make sense of the theory?}
\end{quote}
The answer to this question is not completely obvious and may actually depend on the framework adopted. In the Bondi--Sachs formalism \cite{Madler:2016xju}, which can be viewed as a version of the characteristic initial value problem expressed in terms of an asymptotic expansion near $\scri$, resolution of the equations of motion only requires the specification of the initial value~$q_{ij}|_{u=u_0}$. In that framework it is therefore consistent to view $q_{ij}$ as a dynamical field even though its time evolution $\partial_u q_{ij}=0$ is trivial when the Bondi gauge fixing condition \eqref{Bondi frame} is adopted (more generally the evolution equation is non-trivial \cite{Capone:2021ouo}). The drawback of the Bondi-Sachs formalism is that it requires specification of the shear tensor $C_{ij}$ for all times so that it plays the role of a source rather than that of a dynamical field. Alternatively one can consider the variational principle with boundary conditions at spatial infinity~$i^0$.  As a result one encounters the opposite situation where the shear $C_{ij}$ is free to fluctuate while the sphere metric $q_{ij}$ needs to be fixed as part of the boundary conditions \cite{Mann:2005yr,Compere:2011ve,Virmani:2011gh,Nguyen:2020waf}.\footnote{To be more precise, the only fixed quantity at spatial infinity is the metric $h_{ab}$ on the three-dimensional hyperboloid describing the approach to $i^0$. This metric naturally induces a fixed two-dimensional metric $q_{ij}$ on the celestial sphere $\mathbb{S}^2$.} I believe that these subtleties regarding the choice of formalism and the type of boundary conditions have important implications in selecting one or the other asymptotic symmetry algebra, and therefore deserve further scrutiny. 

\paragraph{The celestial stress tensor.}
The discovery of the extended BMS algebra containing two (centerless) Virasoro subalgebras strongly suggested to rewrite scattering amplitudes in perturbative quantum gravity as correlation functions of a CFT$_2$ living on the celestial sphere. As a result the subleading soft graviton theorem was found equivalent to the familiar conformal Ward identity of a local CFT stress tensor \cite{Kapec:2016jld,Donnay:2022hkf}. This celestial stress tensor is explicitly given by \cite{Kapec:2016jld,Donnay:2021wrk,Donnay:2022hkf}
\begin{equation}
\label{celestial T}
T_{zz}=-\frac{1}{32\pi G} \oint_{\mathcal{C}} \frac{d\zbar}{2i\pi} \int_{-\infty}^\infty du\,  u\, \gamma^2 \left(D^3_z-2 \mathcal{T}_{zz} D_z-D_z \mathcal{T}_{zz}\right)N^{zz}\,, 
\end{equation}
where for ease of the discussion I discarded the dependence on the supertranslation mode. The term in parenthesis can be understood as a derivative operator $\mathcal{D}_z^3$ that transforms covariantly under Weyl rescalings and for which $\mathcal{T}_{zz}$ plays the role of a Weyl connection \cite{Donnay:2021wrk}. By computation of the stress tensor two-point function, the corresponding central charge has further been found to vanish identically \cite{Fotopoulos:2019vac} in agreement with a BMS flux algebra without central extension \cite{Donnay:2021wrk,Donnay:2022hkf}. Because of its Schwarzian transformation, it would have been tempting to identify the quantity $\mathcal{T}_{zz}$ with the celestial stress tensor. However these quantities are very distinct and in particular the celestial stress tensor \eqref{celestial T} does not transform anomalously since the corresponding central charge is zero.

\paragraph{Uplift of the AdS$_3$/CFT$_2$ correspondence.} I come now to my main motivation for reviewing the Schwarzian gauge transformations at null infinity in such details, namely the approach to celestial holography by uplift of the AdS$_3$/CFT$_2$ correspondence \cite{deBoer:2003vf,Cheung:2016iub,Ball:2019atb}. The starting point of this approach is the slicing of Minkowski space\footnote{So far this approach has only been investigated at the level of linearized gravitational perturbations.} by three-dimensional hyperboloids. The slices covering the inner part of the lightcone have negative constant curvature, i.e., they are AdS$_3$ hyperboloids. The key observation is that Virasoro superrotations act tangentially to these slices and further coincide on each slice with the usual Brown--Henneaux asymptotic symmetries \cite{Brown:1986nw}. Since it is widely believed that AdS$_3$ gravity is dual to a  CFT$_2$ -- although the details of this correspondence are still largely mysterious --, the authors of \cite{deBoer:2003vf,Cheung:2016iub,Ball:2019atb} proposed  to leverage the AdS$_3$/CFT$_2$ duality to a correspondence between four-dimensional asymptotically flat gravity and the sought-for celestial CFT. If true this would offer many interesting prospects for understanding quantum gravity in flat space.

The observation that I would like to make here is that this proposal creates some tension regarding the \textit{physical status} of the Schwarzian transformations, and therefore of the value of the Virasoro central charge. It is indeed well-known that the central charge in the AdS$_3$ asymptotic symmetry algebra is nonzero and inversely proportional to the three-dimensional Newton constant (in units of the AdS$_3$ curvature radius) \cite{Brown:1986nw}. The effective three-dimensional Newton constant can be estimated by looking at the four-dimensional Einstein--Hilbert action, yielding \cite{Cheung:2016iub}
\begin{equation}
\label{central charge}
c_{\text{eff}} \sim L_{\text{IR}}^2\, M_{\text{pl}}^2\,,
\end{equation}
where $L_{\text{IR}}$ is an infrared cutoff coming from integration over the direction transverse to the AdS$_3$ slices. This quantity does not appear to vanish in contradistinction with the celestial central charge. Moreover, by explicit coordinate transformation one can show that the uplifted holographic stress tensor \cite{Balasubramanian:1999re} actually maps to $\mathcal{T}_{zz}$ - the $u$-independent component of $\partial_u C_{zz}$ - at null infinity, and both quantities indeed transform anomalously under Virasoro superrotations, i.e, they are associated with a nonzero central charge. Therefore I am forced to conclude that \textit{the uplifted holographic stress tensor is pure gauge and unobservable}.\footnote{The unobservability discussed here refers to the perspective of null infinity $\scri$. It could however play a more physical role for bulk observers such as Rindler observers as suggested in \cite{Pasterski:2022lsl}. I thank Sabrina Pasterski and Herman Verlinde for discussions on this point.} 

Beyond the traditional celestial CFT with zero central charge and stress tensor \eqref{celestial T}, could there be another distinct holographic CFT with nonzero central charge \eqref{central charge} and stress tensor $\mathcal{T}_{zz}$? If it existed such a CFT would apparently describe pure gauge degrees of freedom. The \textit{infrared divergent sector} of flat space scattering amplitudes seems like a natural candidate for what this holographic CFT might govern. Indeed infrared divergences are unphysical and often the result of a poor treatment of gauge invariance, and can be avoided altogether by working with dressed asymptotic states \cite{Kulish:1970ut,Ware:2013zja,Kapec:2017tkm,Arkani-Hamed:2020gyp}. Nonetheless the infrared divergent sector is governed by the correlators of a CFT$_2$ in which the Goldstone modes of spontaneously broken asymptotic symmetries play a prominent role~\cite{Himwich:2020rro,Nguyen:2020waf}. The effective action of supertranslation Goldstone modes already found a natural home in that context, allowing to entirely reconstruct the infrared soft factors from an intrinsically celestial formulation ~\cite{Nguyen:2020waf}. It would be very interesting to assess whether the effective action of superrotation Goldstone modes derived in  \cite{Nguyen:2020hot} -- which clearly describes a CFT$_2$ with stress tensor $\mathcal{T}_{zz}$ -- plays a similar role. This effective action is ubiquitous and similarly appears in the context of the AdS$_3$/CFT$_2$ correspondence \cite{Cotler:2018zff,Nguyen:2020lbg}. From a CFT perspective, it describes the stress tensor sector and can be used to compute Virasoro identity blocks and maximally chaotic out-of-time-order correlators \cite{Cotler:2018zff,Haehl:2018izb,Haehl:2019eae,Anous:2020vtw,Nguyen:2020jqp}. Precisely at the time where these lines are written, an interesting paper has appeared which confirms the existence of a second holographic stress tensor governing chaotic features of asymptotically flat gravity \cite{Pasterski:2022lsl}.   

\section*{Acknowledgments}
I thank Jakob Salzer for past collaboration on related topics. I also thank Romain Ruzziconi for careful reading of the manuscript and useful comments. This work was supported by the STFC grants ST/P000258/1 and ST/T000759/1. 

\appendix

\section{Induced Carrollian connection at null infinity}
\label{app:connection}
I start this appendix by showing that the four-dimensional Levi-Civita derivative operator~$\nabla_\alpha$ induces a torsionfree derivative operator $D_\mu$ at $\scri$ which satisfies
\begin{equation}
\label{properties D}
D_\rho\, q_{\mu\nu}=0\,, \qquad D_\rho\, n^\mu=0\,.
\end{equation}
First we need to split the tangent bundle $T \mM$ at $\scri$ into $T \scri$ and its complement. In particular, we need three basis vectors $e^\alpha_\mu$ labelled by an index $\mu$ that are tangential to $\scri$, i.e., they must satisfy
\begin{equation}
\label{orthogonality}
e^\alpha_\mu\, n_\alpha \hateq 0\,.
\end{equation}
Associated with this basis is a three-dimensional coordinate system $x^\mu$ covering $\scri$ such that
\begin{equation}
e^\alpha_\mu=\frac{\partial x^\alpha}{\partial x^\mu}\,.
\end{equation}
The issue with null hypersurfaces is that there is no canonical splitting between tangential and normal bundles since the normal vector $n^\alpha$ is also tangential. One can still make a choice of splitting, but the induced geometry will depend upon this choice. Fortunately, we are still able to prove that some properties of the induced connection are independent of this choice. The induced covariant derivative of a tangent vector $A^\alpha$ -- which must therefore satisfy $A^\alpha n_\alpha \hateq 0$ -- is defined by projection onto the chosen tangential bundle $T \scri$,
\begin{equation}
D_\rho A^\mu\, \hat{\equiv}\, e^\gamma_\rho\, e^\mu_\alpha\, \nabla_\gamma A^\alpha\,, \qquad A^\mu \,\hat{\equiv}\, e^\mu_\alpha A^\alpha\,,
\end{equation}
where we introduced the dual one-forms $e^\mu_\alpha$ such that $e^\mu_\alpha e^\alpha_\nu=\delta^\mu_\nu$.
Note that the freedom $A^\alpha \to A^\alpha + h\, n^\alpha$ related to the choice of splitting discussed above does not affect this definition thanks to the Bondi condition \eqref{nabla n}. We have
\begin{subequations}
\begin{align}
D_\rho A^\mu&=e^\gamma_\rho\, \nabla_\gamma ( e^\mu_\alpha A^\alpha) -e^\gamma_\rho\, \nabla_\gamma e^\mu_\alpha\, A^\alpha=e^\gamma_\rho\, \partial_\gamma A^\mu -e^\gamma_\rho e^\alpha_\nu\, \nabla_\gamma e^\mu_\alpha\,  A^\nu\\
&\equiv \partial_\rho A^\mu +\Gamma^\mu_{\rho \nu} A^\nu\,,
\end{align}
\end{subequations}
such that we identify the induced connection
\begin{equation}
\label{induced connection}
\Gamma^\mu_{\rho \nu} \hateq -e^\gamma_\rho e^\alpha_\nu\, \nabla_\gamma e^\mu_\alpha\,.
\end{equation}
Using
\begin{equation}
\nabla_\gamma e^\mu_\alpha=\partial_\gamma e^\mu_\alpha-\Gamma_{\gamma \alpha}^\beta e^\mu_\beta=\nabla_\alpha e^\mu_\gamma\,,
\end{equation}
it is straightforward to show that the induced connection is torsionfree, $\Gamma^\rho_{[\mu\nu]}=0$. 

Projection of \eqref{nabla n} directly implies $D_\rho n^\mu=0$, and we are left to show that the induced connection is compatible with the induced metric $q_{\mu\nu}$. For this we first introduce a decomposition of the four-dimensional metric $g_{\alpha \beta}$ in a neighborhood of $\scri$,
\begin{equation}
g_{\alpha \beta}=-(n_\alpha m_\beta+n_\beta m_\alpha)+q_{\alpha \beta}\,,
\end{equation}
where $m^\alpha$ is any null vector normalized such that $m^\alpha n_\alpha=-1$. We then proceed to compute
\begin{equation}
D_\rho q_{\mu\nu}\hateq e^\gamma_\rho e^\alpha_\mu e^\beta_\nu\, \nabla_\gamma q_{\alpha \beta}=e^\gamma_\rho e^\alpha_\mu e^\beta_\nu\, \nabla_\gamma \left(g_{\alpha \beta}+n_\alpha m_\beta+n_\beta m_\alpha \right) \hateq 0\,,
\end{equation}
where in the last line I made use of the Bondi condition \eqref{nabla n} together with \eqref{orthogonality}. As a direct consequence we also have
\begin{equation}
\label{lie q}
\lied_n q_{\mu\nu}=n^\rho D_\rho q_{\mu\nu}+q_{\mu \rho} D_\nu n^\rho+q_{\nu\rho} D_\mu n^\rho=0\,,
\end{equation}
which is simply the projection of the second equation in \eqref{nabla n}. It means that $q_{\mu \nu}$ is the lift along the null direction generated by $n^\mu$ of a two-dimensional non-degenerate euclidean metric on a `cut' of $\scri$ with normal vector $n^\mu$. This concludes the proof of the universal properties \eqref{properties D} satisfied by the induced derivative operator $D_\mu$ at $\scri$. 

For a non-degenerate metric, only the Levi-Civita connection is both metric compatible and torsionfree. For a degenerate metric $q_{\mu\nu}$ however, the conditions \eqref{properties D} are not sufficient to fully determine the connection. From the viewpoint of the four-dimensional bulk geometry, this is partly related to the arbitrariness in splitting the tangent bundle $T \mM$ into $T \scri$ and its complement. More interestingly perhaps, the undetermined connection coefficients also encode non-universal data (namely the shear tensor). The constraints on $\Gamma^\rho_{\mu\nu}$ following from \eqref{properties D} are given for example in appendix~A of \cite{Nguyen:2020hot},
\begin{align}
\label{carrollian Christoffels}
\Gamma^\sigma_{\mu\nu}\, q_{\sigma \rho}=\frac{1}{2}\left(\partial_\mu q_{\nu \rho}+\partial_\nu q_{\mu \rho}-\partial_\rho q_{\mu\nu} \right)\,, \qquad \Gamma^\rho_{\mu\nu}\, n^\nu=0\,.
\end{align}
To make these constraints more explicit, we can introduce a set of basis vectors $(n^\mu, e^\mu_i)$ with $i=1,2$ together with the corresponding coordinate system $(u,x^i)$ that satisfies
\begin{equation}
n^\mu=\frac{\partial x^\mu}{\partial u}\,, \qquad e^\mu_i=\frac{\partial x^\mu}{\partial x^i}\,.
\end{equation}
Again there is no canonical way to split the null direction from the spatial directions, since orthogonality with respect to $q_{\mu\nu}$ is preserved under $e^\mu_i \to e^\mu_i+h_i\, n^\mu$.  Nonetheless $q_{\mu\nu}$ unambiguously projects to a non-degenerate two-dimensional euclidean metric
\begin{equation}
q_{ij}=e^\mu_i  e^\nu_i q_{\mu \nu}\,, \qquad \partial_u q_{ij}=0\,,
\end{equation}
whose time-independence follows from \eqref{lie q}.
The relations \eqref{carrollian Christoffels} thus become
\begin{equation}
\Gamma^k_{ij}=\frac{1}{2} q^{kl} \left(\partial_i q_{jl}+\partial_j q_{il}-\partial_l q_{ij}\right)\,, \qquad \Gamma^i_{uu}=\Gamma^i_{uj}=\Gamma^u_{uu}=\Gamma^u_{ui}=0\,.
\end{equation}
The purely spatial components are the Levi-Civita coefficients associated with the spatial metric $q_{ij}$ induced on $\mathbb{S}^2$. All time components are zero, except $\Gamma^u_{ij}$ which is left completely undetermined. At null infinity $\Gamma^u_{ij}$ is closely related to the shear tensor $C_{ij}$ as explicitly shown in \eqref{Gamma_u}. It therefore encodes non-universal information about gravitational radiation in asymptotically flat spacetimes. 

As a final but important remark, we now explicitly see that the quantity
\begin{equation}
D_\mu A_\nu\,, \qquad \forall A_\mu \quad \text{s.t.} \quad A_\mu n^\mu=0\,,
\end{equation}
does not involve any of the undetermined connection coefficients. Thus the covariant derivative of covectors orthogonal to $n^\mu$ is free of any ambiguity or indeterminacy, and simply coincides with the lift of the covariant derivative on the two-dimensional base space~$\mathbb{S}^2$. A  formula which makes this property explicit was given by Geroch \cite{Geroch1977},
\begin{equation}
D_\mu A_\nu=\partial_{[\mu} A_{\nu]}+\frac{1}{2} \lied_B\, q_{\mu\nu}\,,
\end{equation}
where $B^\mu$ is any vector field satisfying $B^\mu q_{\mu\nu}=A_\nu$.

\bibliography{bibl}
\bibliographystyle{JHEP}
\end{document}